\documentclass[12pt,preprint]{aastex}
\def\Ep{E_{\rm p}}

\def\F0{F_{\rm 0}}
\def\t0{t_{\rm 0}}

\newcommand{\ltsima} {$\; \buildrel < \over \sim \;$}
\newcommand{\gtsima} {$\; \buildrel > \over \sim \;$}
\newcommand{\lta} {\lower.5ex\hbox{\ltsima}}
\newcommand{\gta} {\lower.5ex\hbox{\gtsima}}
\slugcomment{Date: \today}

\begin{document}

\title{Is Thermal Emission in Gamma-Ray Bursts Ubiquitous?}

\author{Felix Ryde}

\affil{Stockholm Observatory, AlbaNova, SE-106 91 Stockholm,
Sweden}

\begin{abstract}

The prompt emission of gamma-ray bursts has yet defied any simple
explanation, despite the presence of a rich observational material
and great theoretical efforts.  Here we show that all the types of
spectral evolution and spectral shapes that have been observed can
indeed be described with one and the same model, namely  a hybrid
model of a thermal and a non-thermal component. We further show
that the thermal component is the key emission process determining
the spectral evolution. Even though bursts appear to have a
variety of, sometimes complex, spectral evolutions, the behaviors
of the two separate components are remarkably similar for all
bursts, with the temperature describing a broken power-law in
time. The non-thermal component is consistent with emission from a
population of fast cooling electrons emitting optically-thin
synchrotron emission or non-thermal Compton radiation. This
indicates that these behaviors are the fundamental and
characteristic ones for gamma-ray bursts.
\end{abstract}


\section{Introduction}

It was early recognized that the spectra of gamma-ray bursts
(GRBs) have a non-thermal character, with emission over a broad
energy range (e.g. \citet{FM}). This typically indicates emission
from an optically-thin source and an initial proposal for GRBs was
therefore an optically-thin synchrotron model from
shock-accelerated, relativistic electrons (e.g. \citet{katz,
Tavani}). The number density of the radiating electrons is assumed
to be typically a power law as a function of the electron Lorentz
factor $\gamma_{\rm e}$ above a  minimum value, $\gamma_{\rm
min}$, with index $-p$. Such a distribution gives rise to a
power-law photon spectrum with index $\alpha = -2/3$ below a break
energy $E_{\rm p} \propto \gamma _{\rm min} ^2$ and a high-energy
power-law with index $\beta = -(p+1)/2$. However, this model has
difficulties in explaining the observed spectra of GRBs which show
a great variation in $\alpha$ and $\beta$ \citep{preece00}. In
particular, a substantial fraction of them have $\alpha > -2/3$,
which is not possible in the model in its simplest form, since
$\alpha = -2/3$ is the power-law slope of the fundamental
synchrotron function for electrons with an isotropic distribution
of pitch angles \citep{pac}. The problem becomes even more severe
for the case when the cooling time of the electrons is shorter
than the typical dynamic timescale. In the typical setting of GRBs
having a relativistic outflow with a bulk Lorentz factor $\Gamma
\sim 100$, the time scales for synchrotron and inverse Compton
losses are $\sim 10^{-6}$ s \citep{ghis00}, which is much shorter
than both the dynamic time scale $R/2\Gamma^2 c \sim 1 \,\, {\rm
s} \,\,(R / 10^{15}\, \,{\rm cm})$, and the integration time scale
of the recorded data, typically 64 ms to 1 s. In such a case the
low-energy power law should be even softer, with $\alpha = -1.5$
\citep{bussard, g00}, now contradicting a majority of the observed
spectra. The spectra are also observed to evolve dramatically
during the course of a burst, both in $\Ep$, as well as in the
power-law indices, in particular $\alpha$. In approximately 60\%
of all bursts, $\alpha$ varies significantly, mainly by becoming
softer (e.g. \citet{crider97}). Some bursts are found to have
quasi-thermal spectra during the initial phases, before they
become non-thermal \citep{ghirlanda, kaneko, ryde04}.

The peak energy from the above distribution of electrons is given
by $E_{\rm p} = \gamma_{\rm m}^2 B_\perp \Gamma$. In the external
shock model $\gamma_{\rm m}$ and  $B_\perp$ are proportional to
the bulk Lorentz factor, which makes $E_{\rm p}\propto \Gamma^4$,
which poses a problem in explaining the relative narrowness of the
observed distribution of peak energies \citep{preece00}, even
including the X-ray flashes. Similarly, for the internal shock
model $\gamma_{\rm m} \propto \Gamma_{\rm rel}$, the relative
Lorentz factor between the two shells that collide, and $E_{\rm p}
\propto B_{\perp} \Gamma$, expected to give a larger scatter as
well.

A third complication arises in explaining the observed correlation
between the burst's peak-energy and luminosity, also known as the
Amati relation (e.g. \citet{LPM, amati, gamati}); the peak energy
is correlated with the isotropically equivalent energy $E_{\rm p}
\propto  E_{\rm iso}^{0.40\pm0.05}$. For the synchrotron,
internal-shock model one expects  $ E_{\rm p} \propto \Gamma^{-2}
L^{1/2} t_{\rm v}^{-1}$ (e.g. \citet{ZM}), where $t_{\rm v}$ is
the typical variability time scale. This requires that both
$\Gamma$ and $t_{\rm v}$ have to be quite similar for all bursts,
which is difficult to imagine. In addition, assuming a typical $L
\propto \Gamma^2$ (e.g. \citet{KRM}) would even lead to an
anti-correlation (see also \citet{RRL}). Additional assumptions
are needed to explain the positive correlation.

Other variations of the synchrotron or/and inverse Compton model
have been suggested (see e.g. \citet{BaringB,LP00,SP04}), however,
none have been able to describe all aspects of the observations in
a convincing manner. To account for these aspects, I argue that
GRBs, in general, have a strong thermal component, which is
accompanied by a non-thermal component of similar strength.

\section{Spectral Modelling}

Recently, in Ryde (2004) I identified bursts which are dominated
by quasi-thermal emission throughout their duration. The
temperature of the emitting matter exhibits a similar behavior for
all of them, with an initially constant, or weakly decreasing,
temperature ($\sim 100$ keV, power-law index $\sim 0$ to  $-0.3$)
and a distinct break into a faster power-law decay with an index
of approximately $-0.6$ to $-1$. I also suggested that bursts that
are observed to be initially thermal, are similar to these but
have an additional non-thermal component that varies in spectral
slope and grows in relative strength with time. This category of
bursts is illustrated in this paper by GRB980306 (all bursts
discussed here were recorded by the BATSE detector on the {\it
Compton Gamma-ray Observatory}). Spectra from three different
times are shown in the lower-most panels in Figure 1. The model
shown consists of a power law $\propto E^{s}$, representing the
non-thermal emission, combined with a Planck function $\propto
(kT)^2\,  x^2 /(\exp(x)-1)$, where $x=E/kT$, $k$ is Boltzmann's
constant and $T$ is the temperature. It is clear that the relative
strength of the non-thermal component increases with time and that
the index $s$ varies, in this particular burst from $\sim-1.5$ to
$\sim -3$ (see Figure 2). This leads to the apparent softening
below the peak energy. Figure 2 also shows that the temperature of
the black body, for this burst, exhibits a similar evolution like
the purely quasi-thermal bursts discussed by Ryde (2004), with a
distinct break in the cooling curve. There is a total of 10 bursts
that have been discussed in the literature from this category
\citep{ghirlanda,ryde04}.

We will now study the spectra of more typical bursts, bursts which
do not have any exceptionally hard $\alpha$-values, nor have any
conspicuous spectral evolution, and therefore a thermal component
is not required in a first appraisal. For this purpose we analyze
the sample of the 25 strongest pulses in the catalogue of
\citet{KRL}, which comprise a complete sample of pulses with a
varying spectral shapes and evolution. We compare the results of
the fits of the hybrid model to those of the most commonly used
\citet{band93} model, which is an empirical function consisting of
a low-energy power-law with index $\alpha$, exponentially
connected to a high-energy power-law with index $\beta$ at an
energy $E _{\rm p}$. We note that these two models have the same
number of parameters; $kT$, $s$, and two amplitudes, compared to
$\alpha$, $\beta$, $E_{\rm p}$, and one amplitude.  The reduced
$\chi^2$ values and the residuals of the fits indicate equally
good fits for both models; the $\chi^2$-values are in most cases
indistinguishable statistically. The hybrid model was formally
better (had a lower $\chi^2_\nu$ value) in 10 of the cases. The
largest differences were for GRB950211 $(\chi^2_{\rm hyb};
\chi^2_{\rm band}) = (1.03;1.10)$ for 540 degrees of freedom
(d.o.f.) and GRB960530 $(\chi^2_{\rm hyb}; \chi^2_{\rm band}) =
(0.975; 0.999)$ for 2071 d.o.f. and finally for GRB950818
$(\chi^2_{\rm hyb}; \chi^2_{\rm band}) = (1.09; 0.96)$ for 1819
d.o.f. If a hybrid model with a sharply broken power law with say
$\alpha  \equiv -1.5$ and $\beta \equiv -2.1$ (motivated in \S 3)
is used instead, the $\chi^2_\nu$ of the latter fit becomes lower:
$1.02$. This illustrates the obvious fact that the simple
power-law model is an approximation of the actual non-thermal
emission if the break energy is within the studied window for a
significant fraction of the burst duration. In comparing the two
models it should also be noted that the hybrid model is a physical
model rather than an empirical model and that the fit results are
reasonable from a theoretical point-of-view (see \S 3). Figure 2
shows three of the studied bursts; GRB921207, GRB950624, which
illustrate the most common behavior in which $s$ evolves from
$-1.5$ to $\sim - 2.1$, and GRB960530 for which $s$ is consistent
of being constant $\sim -1.5$ even though a weak hardening is
indicated. For all the cases the temperature again has a distinct
break in its evolution. Three spectra from GRB950624 are also
shown in Figure 1, illustrating the non-thermal character of the
summed spectrum through out the pulse.

In conclusion, the spectral evolution is very similar from burst
to burst and is independent of the relative strength of the
thermal component. This is in stark contrast to the variety of
apparent spectral behaviors found by using the Band function, for
instance, with strong variation in $\alpha$. This fact is a strong
indication that the thermal emission, combined with a non-thermal
component, is ubiquitous and that the behavior of these components
are the characteristic signatures of GRBs.

\section{Discussion and Conclusion}

It is argued above that thermal radiation is the key feature
during the prompt phase of most GRBs. Apart from the actual fits
presented above and the similarity in behaviors among bursts, such
an interpretation is attractive for several reasons. First, the
value of the low-energy power-law index, $\alpha$, that would be
found if the Band function were to be used is now only a result of
the relative strength of the thermal component and the slope of
the non-thermal component. If the thermal component is strong
and/or the non-thermal component is hard, the resulting spectrum
will have a hard $\alpha$ and vice versa. This gives a new
interpretation of the observed $\alpha$ distribution which has
been a puzzle. Second, the peak of the spectrum is now determined
by $kT$ and is less sensitive to the bulk Lorentz factor,
motivating the narrow dispersion of peak energies. In fact, if the
photosphere occurs during the acceleration phase it is practically
independent of $\Gamma$. Third, the Amati correlation has a
natural explanation since for a thermal emitter the luminosity and
the temperature are correlated. \citet{RM05} show that, somewhat
depending on the details of the dissipation processes, a positive
correlation close to the observed one arises naturally.

A strong photospheric emission at $\gamma-$ray wave lengths is
predicted in most GRB scenarios, such as in kinetic models
\citep{MR00, DM}, in MHD models \citep{DS}, as well as in Poynting
flux models \citep{Lyu}. In the simplest outflow models the
observer-frame temperature should be constant (independent of
collimation) during the acceleration phase, since the adiabatic
losses are compensated by the acceleration. The typical
temperature is
\begin{equation}
kT_0 = \frac{k}{1+z} \left( \frac{L_0}{4 \pi  r_0^2 c a}
\right)^{1/4} \sim 100 \,\,{\rm keV}
\end{equation}
for $L_0 = 10^{51}$ erg, $r_0 =10^8$ cm and $z=1$. After
saturation (the free energy of the outflow has been transferred to
 kinetic energy) the temperature will follow a simple adiabatic
relation, during which the outflow coasts along with a constant
$\Gamma = L_0 / \dot{ M } c^2$, where $L_0$ and $\dot{M}$ are the
luminosity and mass outflow rates. The observed emission from an
optically-thick shell that expands outwards would emit according
to this type of a pattern, similarly to the observed temperature
drops in Figure 2. The timescale for the saturation, which
according to the observed pulses should be around 1 s, leads to
the necessity of very under-loaded fireballs ($\sim 10^{-9}$
M$_{\odot}$). However, the radiative efficiency is, by necessity,
very low for such a scenario, due to large optical depths and that
most energy is in kinetic form. On the other hand, if the outflow
is indeed radiation-dominated then the saturation will naturally
occur at the photosphere. This is also the typical case for
electromagnetic outflows \citep{DS}. Furthermore, Rees \&
M\'esz\'aros (2005) argued that dissipation processes (magnetic
reconnections, shocks) below the photosphere could radically
enhance the thermal luminosity and thus the radiative efficiency
(see also \citet{PW}). Comptonization would convert a fraction of
the dissipated kinetic energy back into thermal energy and thus
re-energize the photosphere, giving typical peak energy of
hundreds of keV. But to keep the spectra quasi-thermal during the
evolution, as is observed, there must be sufficient photons
available to keep the spectra close to those of a black-body.
\citet{GW} (see also \citet{MLR}) noted that the photons may still
be coupled to the matter (e$^{\pm}$ or baryons), ensuring a
quasi-thermal distribution, beyond the radius where the optical
depth has become unity. This occurs when the Compton drag time is
shorter than the dynamic time. The flow then saturates when the
decoupling occurs, now at a very low optical depth. The electron
distribution must after this not be perturbed too much from its
thermal distribution to be able to reproduce the observed spectra.

The non-thermal component could be interpreted as the synchrotron
spectrum from a distribution of fast-cooling electrons. The
initial values of  $s \sim - 1.5$ are expected from electrons that
are cooled to energies below the $\gamma_{\rm min}$ of the
injected electrons. The change in index to $\sim - 2$ could
indicate that the frequency corresponding to $\gamma_{\rm min}$
now moves though the observed energy range and that we, at late
times, are detecting the high-energy power-law of the cooling
spectrum with $ s = -(p+2)/2$. For instance, for a Fermi type of
particle acceleration in relativistic shocks $p \sim 2.2$ and $s =
-2.1$ \citep{mesrev}.

In Figure 3, we plot the energy-flux light-curves of three of the
bursts studied. The relative strengths of the thermal component
vary substantially among them, with GRB980306 having a strong
thermal, initial phase. The obvious correlation between the
thermal and the non-thermal components is noteworthy, indicating
that the emissions cannot be completely independent. Rees \&
M\'esz\'aros (2005) suggested that  the non-thermal emission,
which is superimposed on the thermal Compton-spectrum, is due to
synchrotron shock-emission further out from the photosphere.
Internal outflow variations, leading to the internal shocks, would
be accompanied by corresponding variations in the thermal
emission. An alternative scenario is that non-thermal electrons,
accelerated at a shock close to the photosphere, are cooled
quickly by the thermal radiation, thereby emitting a non-thermal
Compton radiation, boosting every photon by a factor of
$\gamma^2$. The radiation energy-density could then be comparable
or larger than the magnetic energy density. An increase in thermal
emission and energy density from the photosphere would lead to an
increase in the Compton cooling and emission from the non-thermal
electrons. This would naturally explain the close correlation
between the components and that they occur approximately
simultaneously. The variation in $s$ from $\sim -1.5$ to $\sim -2$
would then be interpreted as $\nu_{\min}$ approaching $kT$.

The temperature is shown above to decay as $T \propto t^{-\kappa}
\propto R^{-\kappa}$ during the coasting phase. The thermal energy
flux goes as $F_{\rm BB} \propto A T^4$, where $A$ is the emitting
surface. Due to relativistic abberation of light, the surface
visible to an observer at infinity is $A = \pi R^2/\Gamma^2$.
Therefore, $F_{\rm BB} \propto T^{4-2/\kappa}$ and typical values
of $\kappa = [0.67,1.5]$  give $F_{\rm BB} \propto T^\eta
\,\,\,\,\, {\rm with} \,\, \eta = [1.0,2.7].$ It was further noted
above that the total measured flux is approximately proportional
to the thermal flux component, since the two components track each
other. Therefore, the last relation reproduces the power-law
hardness-intensity correlation, which pulses commonly exhibit, and
the distribution of its power-law indices which was determined to
be $1.9\pm 0.7$ by \citet{BR01}.

\begin{acknowledgements}
I wish to thank Drs. C. Bj\"ornsson, C. Fransson, P. M\'esz\'aros,
and M. Rees for useful discussions. I also thank the referee for
constructive criticism and Karin Ryde for assistance with the
language. Support for this work was given by the Swedish Research
Council.
\end{acknowledgements}

\clearpage

\begin{figure}[]
\epsscale{1}
 \plotone{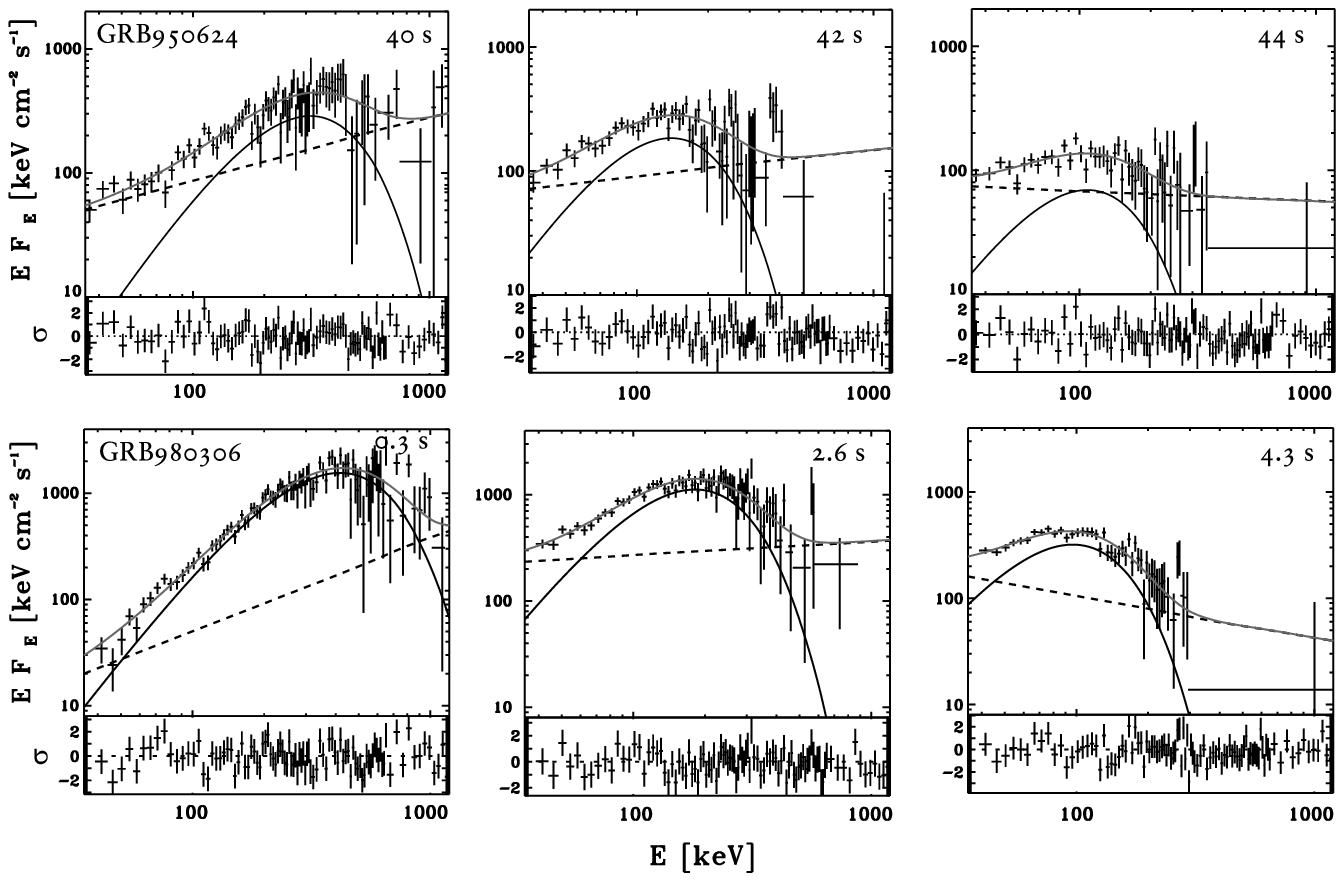}
\caption{Hybrid-model spectral-fits to {\it CGRO} BATSE data
($\sim 20 - 1900$ keV) for two bursts discussed in the text
(GRB950624, GRB980306). The time after the trigger of the fitted
bin is given in the upper right-hand corner. Note that the
investigated pulse in GRB950624 started at 39.8 s. The non-thermal
components is represented by the dashed line, the thermal
component by the thick line, and the summed spectrum by the grey
line. The spectral data points are rebinned to achieve a
signal-to-noise ratio of unity.} \label{fig:f1}
\end{figure}

\clearpage

\begin{figure}[]
 \epsscale{1}
\plotone{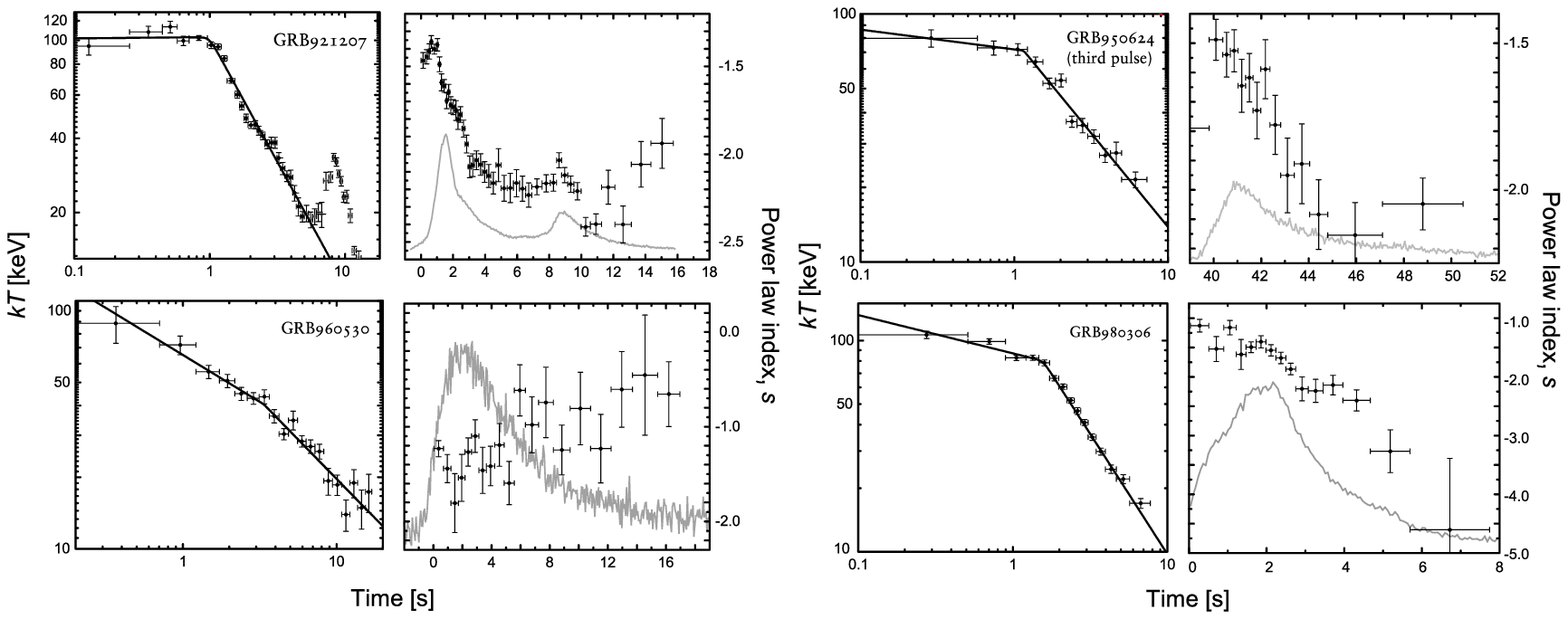} \vspace{-5mm}
 \caption{\small{Evolution of $kT$ (left) and power-law index, $s$ (right).
 The grey curves are the detector count light-curves, with arbitrary normalization.
 The time is counted
 from the trigger, except for in the plot of $kT$ of GRB950624, where time is
 from the beginning of that pulse.}}
 \label{fig:f1}
\end{figure}

\clearpage

\begin{figure}[]
\epsscale{1.0}
 \plotone{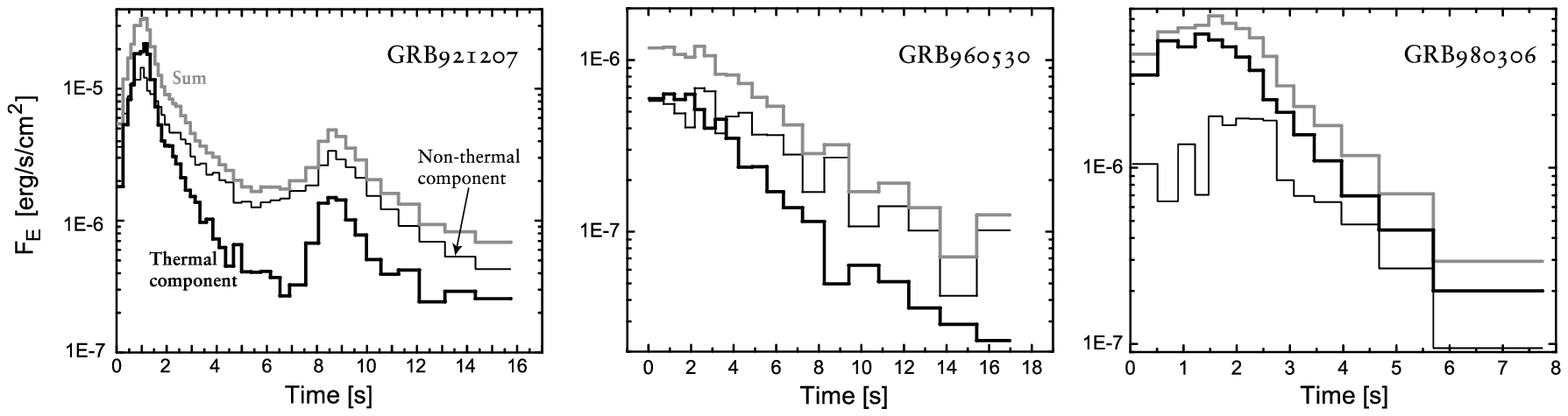}
 \caption{Energy-flux light-curves of the two separate components; thermal (bold line),
 and non-thermal (thin line). The total light-curves are depicted by the grey lines.
 After the break-time the thermal light-curves are approximate power-laws (note the linear time-axis).} \label{fig:f2}
\end{figure}

\end{document}